\documentclass[12pt,preprint]{aastex}

\newcommand{\etal}{{\em et al.}}

\slugcomment{Submitted to the Astrophysical Journal}

\shorttitle{Search for Gamma rays from dSph}
\shortauthors{Acciari et al.}

\begin{document}


\title{VERITAS Search for VHE Gamma-ray Emission from \\
       Dwarf Spheroidal Galaxies}


\author{
V.~A.~Acciari\altaffilmark{1},
T.~Arlen\altaffilmark{2},
T.~Aune\altaffilmark{3},
M.~Beilicke\altaffilmark{4},
W.~Benbow\altaffilmark{1},
D.~Boltuch\altaffilmark{5},
S.~M.~Bradbury\altaffilmark{6},
J.~H.~Buckley\altaffilmark{4},
V.~Bugaev\altaffilmark{4},
K.~Byrum\altaffilmark{7},
A.~Cannon\altaffilmark{8},
A.~Cesarini\altaffilmark{9},
J.~L.~Christiansen\altaffilmark{10},
L.~Ciupik\altaffilmark{11},
W.~Cui\altaffilmark{12},
R.~Dickherber\altaffilmark{4},
C.~Duke\altaffilmark{13},
J.~P.~Finley\altaffilmark{12},
G.~Finnegan\altaffilmark{14},
A.~Furniss\altaffilmark{3},
N.~Galante\altaffilmark{1},
S.~Godambe\altaffilmark{14},
J.~Grube\altaffilmark{11},
R.~Guenette\altaffilmark{15},
G.~Gyuk\altaffilmark{11},
D.~Hanna\altaffilmark{15},
J.~Holder\altaffilmark{5},
C.~M.~Hui\altaffilmark{14},
T.~B.~Humensky\altaffilmark{16},
A.~Imran\altaffilmark{17},
P.~Kaaret\altaffilmark{18},
N.~Karlsson\altaffilmark{11},
M.~Kertzman\altaffilmark{19},
D.~Kieda\altaffilmark{14},
A.~Konopelko\altaffilmark{20},
H.~Krawczynski\altaffilmark{4},
F.~Krennrich\altaffilmark{17},
G.~Maier\altaffilmark{15,\amalg},
S.~McArthur\altaffilmark{4},
A.~McCann\altaffilmark{15},
M.~McCutcheon\altaffilmark{15},
P.~Moriarty\altaffilmark{21},
R.~A.~Ong\altaffilmark{2},
A.~N.~Otte\altaffilmark{3},
D.~Pandel\altaffilmark{18},
J.~S.~Perkins\altaffilmark{1},
M.~Pohl\altaffilmark{17,\mho},
J.~Quinn\altaffilmark{8},
K.~Ragan\altaffilmark{15},
L.~C.~Reyes\altaffilmark{22},
P.~T.~Reynolds\altaffilmark{23},
E.~Roache\altaffilmark{1},
H.~J.~Rose\altaffilmark{6},
M.~Schroedter\altaffilmark{17},
G.~H.~Sembroski\altaffilmark{12},
G.~Demet~Senturk\altaffilmark{24},
A.~W.~Smith\altaffilmark{7},
D.~Steele\altaffilmark{11,\diamond},
S.~P.~Swordy\altaffilmark{16},
G.~Te\v{s}i\'{c}\altaffilmark{15},
M.~Theiling\altaffilmark{1},
S.~Thibadeau\altaffilmark{4},
A.~Varlotta\altaffilmark{12},
V.~V.~Vassiliev\altaffilmark{2},
S.~Vincent\altaffilmark{14},
R.~G.~Wagner\altaffilmark{7},
S.~P.~Wakely\altaffilmark{16},
J.~E.~Ward\altaffilmark{8},
T.~C.~Weekes\altaffilmark{1},
A.~Weinstein\altaffilmark{2},
T.~Weisgarber\altaffilmark{16},
D.~A.~Williams\altaffilmark{3},
S.~Wissel\altaffilmark{16},
M.~D.~Wood\altaffilmark{2},
B.~Zitzer\altaffilmark{12}
}
 
\email{rgwcdf@hep.anl.gov}


\altaffiltext{1}{Fred Lawrence Whipple Observatory, Harvard-Smithsonian Center for Astrophysics, Amado, AZ 85645, USA}
\altaffiltext{2}{Department of Physics and Astronomy, University of California, Los Angeles, CA 90095, USA}
\altaffiltext{3}{Santa Cruz Institute for Particle Physics and Department of Physics, University of California, Santa Cruz, CA 95064, USA}
\altaffiltext{4}{Department of Physics, Washington University, St. Louis, MO 63130, USA}
\altaffiltext{5}{Department of Physics and Astronomy and the Bartol Research Institute, University of Delaware, Newark, DE 19716, USA}
\altaffiltext{6}{School of Physics and Astronomy, University of Leeds, Leeds, LS2 9JT, UK}
\altaffiltext{7}{Argonne National Laboratory, 9700 S. Cass Avenue, Argonne, IL 60439, USA}
\altaffiltext{8}{School of Physics, University College Dublin, Belfield, Dublin 4, Ireland}
\altaffiltext{9}{School of Physics, National University of Ireland Galway, University Road, Galway, Ireland}
\altaffiltext{10}{Physics Department, California Polytechnic State University, San Luis Obispo, CA 94307, USA}
\altaffiltext{11}{Astronomy Department, Adler Planetarium and Astronomy Museum, Chicago, IL 60605, USA}
\altaffiltext{12}{Department of Physics, Purdue University, West Lafayette, IN 47907, USA }
\altaffiltext{13}{Department of Physics, Grinnell College, Grinnell, IA 50112-1690, USA}
\altaffiltext{14}{Department of Physics and Astronomy, University of Utah, Salt Lake City, UT 84112, USA}
\altaffiltext{15}{Physics Department, McGill University, Montreal, QC H3A 2T8, Canada}
\altaffiltext{16}{Enrico Fermi Institute, University of Chicago, Chicago, IL 60637, USA}
\altaffiltext{17}{Department of Physics and Astronomy, Iowa State University, Ames, IA 50011, USA}
\altaffiltext{18}{Department of Physics and Astronomy, University of Iowa, Van Allen Hall, Iowa City, IA 52242, USA}
\altaffiltext{19}{Department of Physics and Astronomy, DePauw University, Greencastle, IN 46135-0037, USA}
\altaffiltext{20}{Department of Physics, Pittsburg State University, 1701 South Broadway, Pittsburg, KS 66762, USA}
\altaffiltext{21}{Department of Life and Physical Sciences, Galway-Mayo Institute of Technology, Dublin Road, Galway, Ireland}
\altaffiltext{22}{Kavli Institute for Cosmological Physics, University of Chicago, Chicago, IL 60637, USA}
\altaffiltext{23}{Department of Applied Physics and Instrumentation, Cork Institute of Technology, Bishopstown, Cork, Ireland}
\altaffiltext{24}{Columbia Astrophysics Laboratory, Columbia University, New York, NY 10027, USA}

\altaffiltext{$\amalg$}{Now at DESY, Platanenallee 6, 15738 Zeuthen, Germany}
\altaffiltext{$\mho$}{Now at Institut f\"{u}r Physik und Astronomie, Universit\"{a}t Potsdam, 14476 Potsdam-Golm,Germany; DESY, Platanenallee 6, 15738 Zeuthen, Germany}
\altaffiltext{$\diamond$}{Now at Los Alamos National Laboratory, MS H803, Los Alamos, NM 87545}


\begin{abstract}
  Indirect dark matter searches with ground-based gamma-ray
  observatories provide an alternative for identifying the particle
  nature of dark matter that is complementary to that of direct search
  or accelerator production experiments.  We present the results of
  observations of the dwarf spheroidal galaxies Draco, Ursa
  Minor, Bo\"{o}tes 1, and Willman 1 conducted by the Very Energetic
  Radiation Imaging Telescope Array System (VERITAS).  These galaxies are
  nearby dark matter dominated objects located at a typical distance of
  several tens of kiloparsecs for which there are good measurements of the dark
  matter density profile from stellar velocity measurements.  Since the 
  conventional astrophysical background of very high energy gamma rays
  from these objects appears to be negligible, they are good targets to search
  for the secondary gamma-ray photons produced by interacting or decaying
  dark matter particles.  No significant gamma-ray flux above 200 GeV was
  detected from these four dwarf galaxies for a typical exposure of $\sim20$
  hours.  The 95\% confidence upper limits on the integral gamma-ray flux are
  in the range $0.4-2.2\times10^{-12}~\mbox{photons~cm}^{-2}\mbox{s}^{-1}$.
  We interpret this limiting flux in the context of pair annihilation of
  weakly interacting massive particles (WIMPs) and derive constraints on
  the thermally averaged product of the total self-annihilation cross section
  and the relative velocity of the WIMPs
  ($\langle\sigma v\rangle \lesssim 10^{-23}$ cm$^3$ s$^{-1}$ for
  $m_\chi \gtrsim 300~\mbox{GeV}/\mbox{c}^2$).  This limit is obtained under conservative
  assumptions regarding the dark matter distribution in dwarf galaxies and
  is approximately three orders of magnitude above the generic theoretical
  prediction for WIMPs in the minimal supersymmetric standard model
  framework.  However significant uncertainty exists in the dark matter
  distribution as well as the neutralino cross sections which under
  favorable assumptions could further lower this limit.
\end{abstract}




\section{Introduction}     \label{sec:intro}

The existence of astrophysical non-baryonic dark matter (DM) has been
established by its gravitational effects on a wide range of spatial
scales.  Perhaps the most compelling
evidence for the existence of weakly interacting particle dark matter
comes from observations of colliding galaxy clusters in which the baryonic
matter in the form of X-ray emitting gas is separated from the source of
the gravitational potential detected through gravitational lensing \citep{clo06,bra08}.
However, despite the well-established presence of DM in the universe,
its particle nature is unknown.

The quest to understand the nature of DM draws upon research in cosmology,
particle physics, and astroparticle physics with direct and indirect detection
experiments \citep{ber00,bertone05}.
In this paper we focus on the indirect search for very high energy (VHE, Energy $>$ 100 GeV)
gamma rays resulting from the interaction or decay of DM particles in
astrophysical objects in which the gravitational potential is
dominated by DM.

Among many theoretical candidates for the DM particle \citep{tao08}, a
weakly interacting massive particle (WIMP) is among the best
motivated.  A thermal relic of the early universe with an interaction
cross section on the weak scale will naturally produce the present-day
DM density if the particle has a weak-scale
mass \citep{lee77,dic77}($\Omega_{DM}h^2 = 0.1099 \pm 0.0062$ (WMAP only),
$\Omega_{DM}h^2 = 0.1131 \pm 0.0034$ (WMAP + Baryon Acoustic Oscillations + Type Ia Supernovae),
where $\Omega_{DM}$ is the ratio
of dark matter density to the critical density for a flat universe and
$h$ is a dimensionless quantity defined as the Hubble constant, $H_\circ$,
normalized to $100 \mbox{km~s}^{-1}\mbox{Mpc}^{-1}$ \citep{kom09}).
Several candidates for WIMPs are predicted
in extensions to the standard model of particle physics, for example,
the neutralino from supersymmetry \citep{ell84} and the Kaluza-Klein
particle in theories of universal extra dimensions
\citep{servant03,bertone03}.  Both neutralinos and Kaluza-Klein particles are
predicted to have a mass in the range of a few tens of GeV/c$^2$ to possibly
a few TeV/c$^2$.

The self-annihilation of WIMPs produces a unique spectral signature of
secondary gamma rays which is expected to significantly deviate from
the standard power-law behavior observed in most conventional astrophysical
sources of VHE gamma rays and would have a cutoff at the WIMP mass.
In addition, it could exhibit a monoenergetic line at the WIMP mass or
a considerable enhancement of gamma-ray photons at the endpoint of the
spectrum due to the internal bremsstrahlung effect \citep{bri08}.
Observation of these spectral signatures combined with the spatial distribution
of the gamma-ray flux from an astrophysical source is a unique capability of indirect
DM searches utilizing gamma rays.

Nearby astrophysical objects with the highest dark matter density
are natural candidates for indirect DM searches.  While the Galactic Center
is likely to be the brightest source of annihilation radiation (e.g.\ see
\citealt{ber98}), VHE gamma-ray measurements reveal a bright gamma-ray source at the
center which constitutes a large astrophysical background \citep{aha06a}.
Other possible bright sources are expected to be the cores of nearby large galaxies such as
M31 or halos around galactic intermediate mass black holes, should they exist,
where adiabatic compression of dark matter halos could result
in a large enhancement in the annihilation signal, in some cases already
exceeding experimental bounds \citep{bri09a,ber09}.
However, in these regions, the DM density profiles are
poorly constrained and, in the case of nearby galaxies,
conventional astrophysical VHE sources can generate 
backgrounds for DM annihilation searches.  In contrast, the satellite dwarf
spheroidal galaxies (dSphs)
of the Milky Way are attractive targets for indirect dark
matter searches due to their proximity (20-100 kpc) and relatively
well-constrained DM profiles derived from stellar kinematics.  They, in fact,
may be the brightest sources for annihilation radiation after the Galactic Center \citep{bul09}.
The general lack of active or even recent star formation in most dSphs
implies that there is little background from conventional
astrophysical VHE processes as has been observed in the Milky Way
Galactic Center \citep{kos04,aha06b,aha09}.  The growing class of nearby dSphs
discovered by recent all-sky surveys \citep{york00, belokurov07} increases the probability
of finding an object for which the halo density is sufficient to yield a detectable
gamma-ray signal.

In this paper, we report on an indirect DM search for gamma rays from
four dSphs: Draco, Ursa Minor, Bo\"{o}tes 1 and Willman 1, carried out
using the Very Energetic Radiation Imaging Telescope Array System
(VERITAS).  After a brief summary of the properties of the
observational targets and previous VHE observations in Section 2, we
describe the VERITAS instrument, the data set, and the analysis
techniques in Section 3.  Sections 4 and 5 are devoted to the
discussion of the results and their interpretation in terms of
constraints on the WIMP parameter space.  We conclude in Section 6
with a discussion of the opportunities for indirect DM detection by
future ground-based gamma-ray instrumentation.

\section{Observational Targets}     \label{sec:targets}

\begin{table}
\begin{center}
  \caption{Properties of the four dSphs.  Preferred values for
    DM halo parameters, $\rho_s$ and r$_s$, which are defined in the text
    are taken from \citet{str07}
    and \citet{bri09b}.  Values for L$_{V}$, the visual luminosity, and
    r$_{h}$, the half-light radius are taken from
    \citet{wal09}.  $R_d$ is heliocentric distance of the dSph.  The calculation
    of the dimensionless line of
    sight integral, J, which is normalized to the critical density
    squared times the Hubble radius ($3.832\times10^{17}
    \mbox{GeV}^2\mbox{cm}^{-5}$), is explained in Section
    \ref{sec:limits}.  The J value for Bo\"{o}tes was calculated by G.D. Martinez
    and J.S. Bullock.
    As explained in the text, the elongation of Bo\"{o}tes and the relative lack
    of stellar kinematic data lead to large uncertainties for $r_s$ or $\rho_s$
    and no values are provided in this case. \label{tbl:J}}

\begin{tabular}{lcccc}
\tableline\tableline
Quantity   &  Draco  &  Ursa Minor  &  Bo\"{o}tes 1& Willman 1    \\
\tableline
$\alpha$ [J2000.0]&
17$^h$20$^m$12.4$^s$&
15$^h$ 09$^m$11.3$^s$&
14$^h$00$^m$06$^s$&
10$^h$49$^m$22.3$^s$ \\
$\delta$ [J2000.0]&
57$^\circ$54$'$55$''$&
67$^\circ$12$'$52$''$&
14$^\circ$30$'$00$''$&
51$^\circ$03$'$03$''$\\
L$_{V}$ [L$_\odot$]&$(2.7\pm 0.4)\times 10^5$&$(2.0\pm 0.9)\times 10^5$&$(3.0\pm 0.6)\times 10^4$
&$(1.0\pm 0.7)\times 10^3$\\
r$_{h}$ [pc]&$221\pm 16$&$150\pm 18$&$242\pm 21$&$25\pm 6$  \\
$R_d$ [kpc]&80&66&62&38\\
$\rho_s~[\mbox{M}_\sun/\mbox{kpc}^3]$   &  
$4.5\times 10^7$&$4.5\times 10^7$&---&$4\times10^8$  \\
$r_s$ [kpc]&  
0.79&0.79&---&0.18           \\
$J(\rho_s,r_s)$&  4  &  7  &  3 & 22        \\
\tableline
\end{tabular}
\end{center}
\end{table}

Three of the dSphs forming the subject of this paper, Draco, Ursa
Minor, and Willman 1, have been identified as the objects within the
dSph class with potentially the highest gamma-ray self-annihilation
flux, e.g. see \citet{str07,str08}.  The modeling of the DM distribution
of these galaxies usually is based on stellar kinematics assuming a spherically symmetric
stellar population and an NFW profile for DM \citep{nav97}
characterized by two parameters: the scale radius $r_{s}$ and scale
density $\rho_{s}$,
\begin{equation}\label{eqn:nfw_profile}
\rho(r) = \rho_{s}\left(\frac{r}{r_s}\right)^{-1}
\left(1+\frac{r}{r_s}\right)^{-2}.
\end{equation}
The properties of these galaxies including constraints on $r_{s}$ and
$\rho_{s}$ as found in \citet{str07} and \citet{str08} are summarized
in Table \ref{tbl:J}.  

The Draco dSph is one of the most frequently studied objects for
indirect DM detection
\citep{bal00,tyl02,eva04,col07,san07,str07,str08,bri09b}.  It has an approximately
spherically symmetric stellar distribution \citep{irw95} with total luminosity
of the order of $10^5$ L$_\odot$ \citep{piatek02}.  The large spectroscopic data set
available for this object \citep{wilkinson04,mun05,wal07} tightly
constrains its DM distribution profile.  Draco is consistent with an
old low-metallicity ([Fe/H] = -1.8 $\pm$ 0.2) stellar population with
no significant star formation over the last 2 Gyrs \citep{apa01}.
Draco previously has been observed at VHE energies by the STACEE
observatory \citep{dri08}, the Whipple 10m telescope \citep{woo08}, and
the MAGIC telescope \citep{alb08}.

The Ursa Minor dSph has a distance and inferred DM content similar to those of
Draco.  There is no evidence of young or intermediate
age stellar populations in Ursa Minor \citep{shetrone01}.  Photometric
studies of this object have found evidence for significant structures
in the stellar distribution in the central $10'$ \citep{bel02,kleyna03}
and an extratidal stellar population \citep{palma03}.  These unusual
morphological characteristics could be evidence of possible
tidal interaction with the Milky Way, velocity projection effects
along the line of sight, or the presence of fluctuations in the DM
induced gravitational potential \citep{kleyna03}.  In fact, such confusing
factors are present in most dSph galaxies.  Ursa Minor was
previously studied at VHE energies by the Whipple 10m telescope
\citep{woo08}.

The recently discovered dSph Bo\"{o}tes 1 \citep{belokurov06} shows
evidence for elongation of the stellar profile.  N-Body
simulations can not reproduce the observed velocity dispersion without
a dominant contribution from DM.  In addition, modeling of the tidal
interaction effects between Bo\"{o}tes 1 and the Milky Way do not
provide an adequate explanation for
the elongation of this system suggesting a non-spherically
symmetric distribution of DM in the Bo\"{o}tes progenitor
\citep{fellhauer08}.  Given the stellar kinematical data is based
on about 30 stars, the scale radius and density of the NFW profile have
large uncertainty as well as signficant degeneracy.  Thus, values for 
$r_{s}$ and $\rho_{s}$ are unavailable in the literature.
The modeling of Bo\"{o}tes 1 was done by
G.D. Martinez and J.S. Bullock (2009, private communication) for a range of NFW fits.
The methodology is described in \citet{martinez09,abd10}.
They produce a probability density function (pdf) for $J$,
the astrophysical contribution to the flux (see Section~\ref{sec:limits}),
which is approximately Gaussian in $\log(J)$.  The value given in
Table~\ref{tbl:J} represents $J$ at the peak of the pdf which is approximately
the mean of the distribution.  The estimates of the age of the stellar
population and metallicity suggest similarity with the old and
metal-poor ([Fe/H] $\sim$ -2.5 -- -2.1) stellar distribution of M92
\citep{belokurov06,munoz06,martin07}.  To date, no other VHE
gamma-ray observations have been reported for this object.

Together with Bo\"{o}tes 1, Willman 1 belongs to the new class of low
surface brightness dSphs recently discovered by the Sloan Digital Sky
Survey \citep{wil05}.  Willman 1 is one of the smallest ($r_h \sim$ 25 pc)
and least luminous ($L_V \sim$ 10$^3$ L$_\odot$) dSphs known.  Its
half-light radius and absolute magnitude suggest that it may be an
intermediate object between dwarf galaxies and globular clusters
\citep{belokurov07}.  Due to the small kinematic sample of stars
available for this object, the constraints on the DM halo
parametrization are poor \citep{str08}.
Latest estimates of the metallicity suggest a low value of [Fe/H]
which is consistent with the observed trend of decreasing metallicity
for fainter dSphs \citep{siegel08}.  The MAGIC collaboration has
recently reported the results of the observations of Willman 1
\citep{ali09}.

\section{Data and Analysis}     \label{sec:obs}

\subsection{The VERITAS Observatory}     \label{sec:veritas}

The VERITAS observatory is an array of four 12m imaging atmospheric
Cherenkov telescopes (IACTs) located at the Fred Lawrence Whipple
Observatory ($31\degr57\arcmin{\rm N}\ 111\degr37\arcmin{\rm W}$) in
southern Arizona at an altitude of 1.27 km above sea level
\citep{wee02}.  The observatory is
sensitive over an energy range of 150 GeV to 30 TeV with an energy
resolution of 10-20\% and an angular resolution (68\% containment) of $< 0.14\degr$ per event.
For the measurements reported here, VERITAS had a point source sensitivity
capable of detecting gamma rays with a flux of 5\% (1\%) of the Crab Nebula
flux above 300 GeV at five standard
deviations in $<2.5$ ($<50$) hours at 20$^\circ$ zenith angle.
During summer, 2009 subsequent to the four dSph observations, the array
configuration was changed, improving the point source sensitivity.
Further technical description of the VERITAS observatory can
be found in \citet{acc08}.

\subsection{Data}

Observations of the Draco, Ursa Minor, Bo\"{o}tes 1, and Willman 1 dSphs
were performed during 2007-2009 (see Table
\ref{tbl:dsph-obs}).  Observations were taken in ``wobble'' mode
\citep{berge07} with the source offset by 0.5$^\circ$ from the center of
the field of view in order to obtain source and background measurement
within the same observation.  The direction of the offset was
alternated between north, south, east, and west to minimize systematic
errors.  {\it Reflected} background regions are defined within the field of view at
the same radius with respect to the camera center as that of the
targeted dwarf galaxy.  Observations were made with varying
atmospheric conditions during moonless periods of the night.  Data were
quality selected for analysis based on the stability of the cosmic-ray
trigger rate and the rms temperature fluctuations observed by an FIR camera
viewing the sky in the vicinty of the observed target ($\leq 0.3\degr C$).
The total exposure on each source is given in Table \ref{tbl:dsph-obs}.

\begin{table}
\begin{center}
  \caption{Summary of observation periods and exposures of
    dSphs by VERITAS.\label{tbl:dsph-obs}}
\begin{tabular}{lccc}
\tableline\tableline
Source       &  Period &  Exposure [hr]& Zenith Angle [$\circ$]\\
\tableline
Draco        &   2007 Apr-May       &    18.38 & 26 -- 51   \\
Ursa Minor   &   2007 Feb-May       &    18.91 & 35 -- 46   \\
Bo\"{o}tes 1 &   2009 Apr-May       &    14.31 & 17 -- 29   \\
Willman 1    &   2007 Dec-2008 Feb  &    13.68 & 19 -- 28   \\
\tableline
\end{tabular}
\end{center}
\end{table}

\subsection{Analysis}     \label{sec:analysis}

Data reduction follows the methods described in \citet{acc08}.  A
brief outline of the analysis flow follows.  Images recorded by
each of the VERITAS telescopes are characterized by a second moment
analysis giving the Hillas parameters \citep{hil85}.  A stereoscopic analysis of
the image parameters is used to reconstruct the gamma ray arrival
direction and shower core position.  The background of cosmic rays is
reduced by a factor of $>10^{5}$ utilizing cuts on the
reconstructed arrival direction ($\theta^2 < 0.013$ deg$^2$) and the image
shape parameters, mean scaled width and length ($0.05 \leq msw \leq
1.16$ and $0.05 \leq msl \leq 1.36$).  The image distance from the
center of the camera is required to be less than $1.43\degr$ to avoid
truncation effects at the edge of the 3.5$^\circ$ field of view.  The
integrated charge recorded in at least two telescopes is further
required to be $>$ 75 photoelectrons (400 digital counts) which
effectively sets the energy threshold of the analysis to be above
$\sim$200 GeV depending on the zenith angle.  The energy threshold
quoted in our analyses is
taken to be the energy at which the differential detection rate
of gamma rays from the Crab Nebula peaks.  The cuts applied
in this analysis were optimized to maximize significance of the
detection for a hypothetical source with a power-law spectrum
(dF/dE = 3.2$\times$ 10$^{-12}$ (E/TeV)$^{-2.5}$ cm$^{-2}$
s$^{-1}$ TeV$^{-1}$) corresponding to 3\% of the Crab Nebula flux.
Two independent data analysis packages were used to analyze the data
and yielded consistent results.

The significance of the detection was calculated by comparing the
counts in the source region to the expected background counts.  The
background in the source region is estimated using the {\it reflected
region} model.  In this model circular background regions, here of
angular radius $0.115\degr$, are defined with an offset from the camera
center equal to that of the putative source.
Eleven background regions can be accommodated
within the VERITAS field of view.  The absence of bright stars within
any of the four dSph pointings allows all eleven regions to be used in the
background count estimation.  The significance of any signal is
computed using the Li and Ma method \citep[eqn. 17]{li83}.

\section{Results}     \label{sec:results}

Table~\ref{tbl:results} summarizes the results for each of the four
dSphs.  The effective energy threshold for each of the targets changes
primarily due to the average zenith angle of observations.  The table
shows the average effective collecting area for gamma rays as calculated
from a sample of simulated gamma-ray showers.
No significant excesses of counts above background were detected from
these observations.  The 95\% confidence level upper limits on the
gamma-ray integral flux were calculated using the bounded profile likelihood
ratio statistic developed by \citet{rol05}.

As we have noted, Draco, Ursa Minor, and Willman I have been observed by other
IACTs and we briefly compare our flux limits to the other observations.  For Draco,
STACEE \citep{dri08} finds a spectral limit of less than
$1.6 \times 10^{-13} \left(\frac{E}{200{\rm GeV}}\right)^{-2.2}\mbox{cm}^{-2} \mbox{s}^{-1}\mbox{GeV}^{-1}$.
The MAGIC flux limit \citep{alb08}
from observation of Draco is $1.1 \times 10^{-11} \mbox{cm}^{-2} \mbox{s}^{-1}$ above a
threshold of 140 GeV.  MAGIC also has set flux limits for Willman 1 in the range
$5.7-9.9 \times 10^{-12}\mbox{cm}^{-2} \mbox{s}^{-1}$ above 100 GeV based on several benchmark
models \citep{ali09}, compared to our limit of $1.17 \times 10^{-12}\mbox{cm}^{-2} \mbox{s}^{-1}$
above a threshold of 320 GeV.  The limits for the VERITAS observations of Draco and Ursa Minor
are an improvement of about a factor of 40 over the earlier observations of the group on
the Whipple 10m IACT \citep{woo08}.

\begin{figure}
\epsscale{0.80}
\plotone{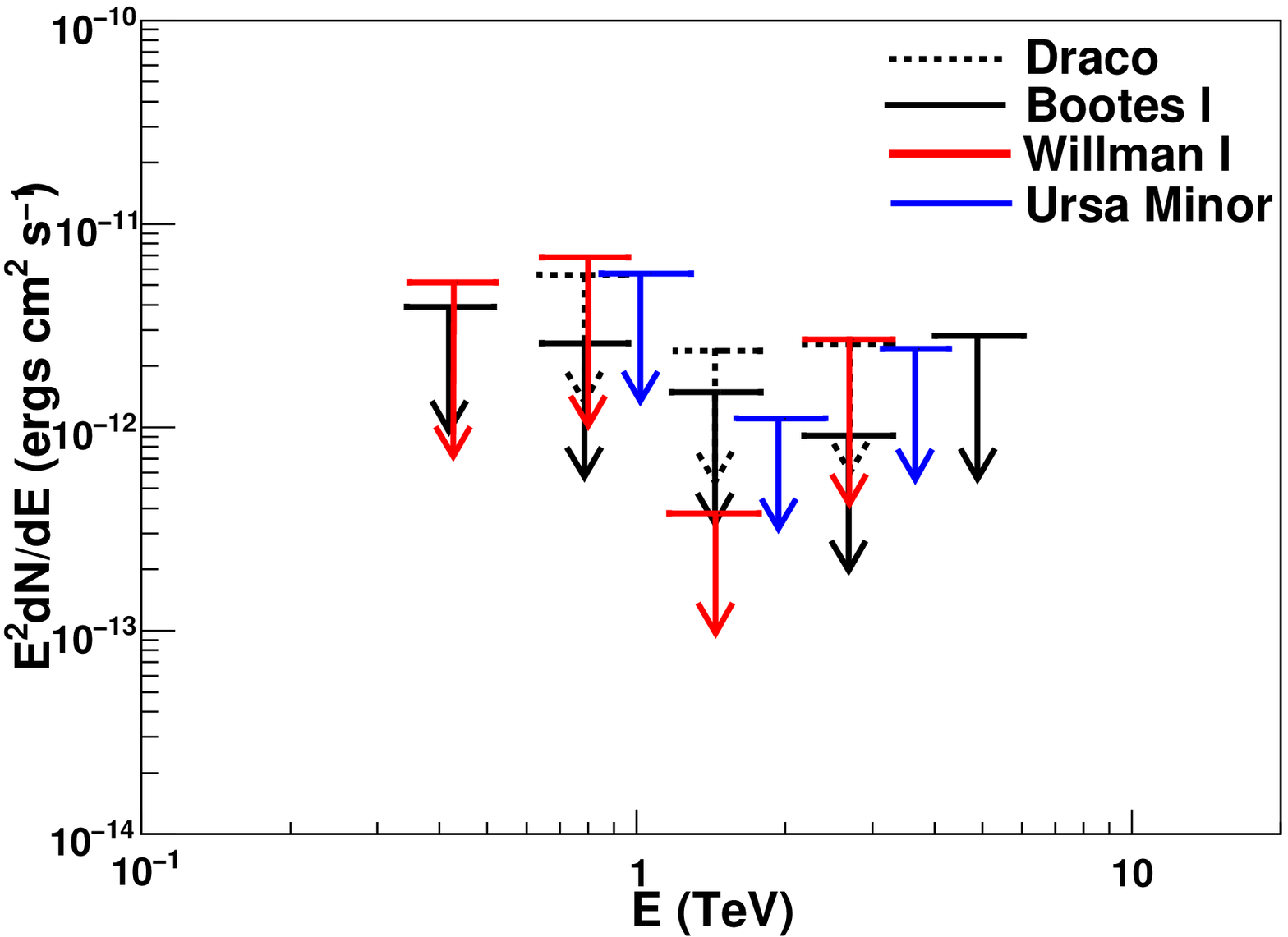}
\caption{95\% C.L. upper limits on the spectral energy density (erg
  cm$^{-2}$ s$^{-1}$) as a function of gamma-ray
  energy for the four dSphs.\label{fig:difflimit-plot}}
\end{figure}

Figure \ref{fig:difflimit-plot} shows the upper limits on the
differential spectral energy density (E$^2$ d$\phi$/dE) as a function of
energy.  The upper limits were derived with four equidistant log
energy bins per decade requiring 95\% C.L. in each bin.

\begin{table}
\begin{center}
\caption{Results of observations of dSphs by VERITAS.\label{tbl:results}}
\begin{tabular}{lcccc}
\tableline\tableline
Quantity                                 &  Draco  &  Ursa Minor  & Bo\"{o}tes 1    & Willman 1\\
\tableline
Exposure [s]                             &  66185  &    68080     & 51532       & 49255    \\
On Source [counts]                       &    305  &      250     &   429       &   326    \\
Total Background [counts]                &   3667  &     3084     &  4405       &  3602     \\
Number of Background Regions             &     11  &       11     & 11          & 11       \\
Significance\tablenotemark{a}            &  -1.51  &    -1.77     & 1.35        & -0.08    \\
95\% C.L. [counts]\tablenotemark{b}      &   18.8  &     15.6     & 72.0        & 36.7     \\
Average Effective Area [cm$^2$]                   &  $5.84\times 10^8$  &  $5.71\times 10^8$    & $6.37\times 10^8$  & $6.37\times 10^8$    \\
Energy Threshold [GeV]\tablenotemark{c}  &    340  &      380     & 300         & 320      \\
Flux Limit 95\% C.L. [cm$^{-2} s^{-1}$]   & $0.49 \times 10^{-12}$  &  $0.40 \times 10^{-12}$ &  $2.19 \times 10^{-12}$ & $1.17 \times 10^{-12}$    \\
\tableline
\end{tabular}
\tablenotetext{a}{\ Li and Ma method \citep{li83}.}
\tablenotetext{b}{Rolke method \citep{rol05}.}
\tablenotetext{c}{Definition given in text.}

\end{center}
\end{table}

\section{Limits on WIMP Parameter Space}     \label{sec:limits}
 
The differential flux of gamma rays from WIMP self-annihilation is given by
\begin{equation}
\frac{d\phi(\Delta\Omega)}{dE} = 
\frac{\langle\sigma v\rangle}{8\pi m_{\chi}^2}
     \left[\frac{dN(E,m_\chi)}{dE}\right]  \int_{\Delta\Omega}d\Omega
     \int \rho^2(\lambda,\Omega)\, d\lambda,
\label{eqn:diff-flux}
\end{equation}
where $\langle\sigma v \rangle$ is the thermally averaged product of
the total self-annihilation cross section and the velocity of the
WIMP, $m_\chi$ is the WIMP mass, $dN(E,m_\chi)/dE$ is the differential
gamma-ray yield per annihilation, $\Delta\Omega$ is the observed solid
angle around the dwarf galaxy center, $\rho$ is the DM mass density,
and $\lambda$ is the line-of-sight distance to the differential
integration volume.  The astrophysical contribution to the flux can be
expressed by the dimensionless factor $J$
\begin{equation}
J(\Delta\Omega) = \left(\frac{1}{\rho_c^2 R_H}\right)
\int_{\Delta\Omega}\, d\Omega\, \int \rho^2(\lambda,\Omega)\, d\lambda,
\end{equation}
which has been normalized to the product of the square of the
critical density, $\rho_c = 9.74\times10^{-30} \mbox{g cm}^{-3}$ and
the Hubble radius, $R_H = 4.16$ Gpc following \citet{woo08}.

\begin{figure}
\epsscale{0.80}
\plotone{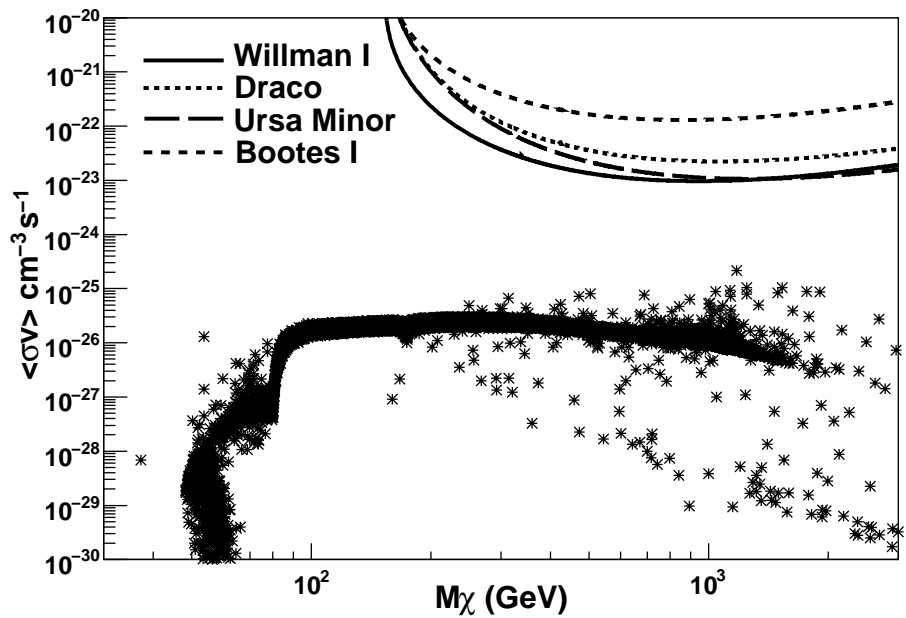}
\caption{Exclusion regions in the $(M_\chi, \langle\sigma v\rangle)$
  parameter space based on the results of the observations.  It is
  computed according to eq.~\ref{eqn:limit} using a composite
  neutralino spectrum (see \citet{woo08}) and the values of $J$ from
  Table~\ref{tbl:J}.  Black asterisks represent points from MSSM
  models that fall within $\pm3$ standard deviations of the relic
  density measured in the 3 year WMAP data set
  \citep{spe07}. \label{fig:limit-plot}}
\end{figure}

Based on equation \ref{eqn:diff-flux}, the upper limits on the
gamma-ray rate, $R_\gamma(\mbox{95\% C.L.})$, constrain the WIMP
parameter space $(m_\chi, \langle\sigma v\rangle)$ according to
\begin{eqnarray}
\frac{R_\gamma(\mbox{95\% C.L.})}{\mbox{hr}^{-1}}  &  >  &
     \frac{J}{1.09 \times 10^{4}}\left(\frac{\langle\sigma v \rangle}
     {3\times10^{-26}\mbox{cm}^3 \mbox{s}^{-1}}\right)           \nonumber     \\
   &    &   \times\int_{0}^\infty \frac{A(E)}{5 \times 10^{8} \mbox{cm}^2}
\left(\frac{\mbox{300 GeV}/\mbox{c}^2}{m_\chi}\right)^2\frac{EdN/dE(E,m_\chi)}{10^{-2}}\frac{dE}{E},
\label{eqn:limit}
\end{eqnarray}
where $A(E)$ is the energy-dependent gamma-ray collecting area.  The
expression has been cast as a product of dimensionless factors with
the variables normalized to representative quantities, e.g. the
cross section times velocity is normalized to
$3\times 10^{-26}\mbox{cm}^3\mbox{s}^{-1}$ which is a rough 
generic prediction for $\left<\sigma v\right>$ for a WIMP thermal
relic in the absence of coannihilations for $m_\chi > 100$ GeV/c$^2$
(c.f. figure \ref{fig:limit-plot}).  The main
contribution to the integral comes from the energy range in the
vicinity of the energy threshold ($E \simeq 300$ GeV for observations
in this paper) where $A(E)$ changes rapidly.  For VERITAS the
effective area at 300 GeV is $\sim6 \times 10^{8}~\mbox{cm}^2$.  For a
representative MSSM model, $EdN/dE$ at 300 GeV is a function of
neutralino mass, $m_\chi$, and it changes in the range
$10^{-2}-10^{-1}$ for $m_\chi$ from 300 GeV/c$^2$ to a few TeV/c$^2$.  Although
$EdN/dE$ is a rapid function of $m_\chi$, this dependence is nearly
compensated by the $(\mbox{300 GeV}/\mbox{c}^2/m_\chi)^2$ prefactor.  The product
of these two contributions and, consequently, the overall integral value,
is weakly dependent on the neutralino mass within the indicated range
and is on the order of 1.  It is evident from the inequality (Equation
\ref{eqn:limit}) that for a typical upper limit on the detection rate
of 1 gamma ray per hour, significantly constraining upper limits on
$\left<\sigma v\right>$ could be established if $J$ is on the order of
$10^{4}$.

Because the factor, $J$, is proportional to the DM density squared, it
is subject to considerable uncertainty in its experimental
determination.  For example, the mass of a DM halo is determined by
the interaction of a galaxy with its neighbors and is concentrated in
the outer regions of the galaxy.  Unlike the DM halo mass, the
neutralino annihilation flux is determined by the inner regions of the
galaxy where the density is the highest.  For these regions the stellar
kinematic data do not fully constrain the DM density profile due to
limited statistics.  Various parametrizations of the DM mass
density profile have been put forward \citep{nav97,kaz04,deb01,bur95}
based on empirical fits and studies of simulated Cold Dark Matter (CDM) halos.
We adopt the assumption of the NFW profile \citep{nav97} given in Equation
\ref{eqn:nfw_profile} which describes a smooth distribution of DM
with a single spatial scale factor $r_{s}$.  The astrophysical factor,
$J$, is then given by
\begin{equation}
J(\Delta\Omega) = \left(\frac{2\pi\rho_s^2}{\rho_c^2 R_H}\right)
\int_{\cos(0.115\degr)}^1\, \int_{\lambda_{min}}^{\lambda_{max}}
\left(\frac{r(\lambda)}{r_s}\right)^{-2}
\left[1+\left(\frac{r(\lambda)}{r_s}\right)\right]^{-4}\,
d\lambda\, d(\cos\theta),
\end{equation}
where the lower integration bound of $0.115\degr$ corresponds to the
size of the signal integration region.  The galactocentric distance,
$r(\lambda)$, is determined by
\begin{equation}
r(\lambda) = \sqrt{\lambda^2 + R_{dSph}^2 - 2\lambda R_{dSph}\cos\theta},
\end{equation}
where $\lambda$ is the line of sight distance and $R_{dSph}$ is the distance
of the dwarf galaxy from the Earth.  

Although the integration limits, $\lambda_{min}$ and $\lambda_{max}$,
are determined by the tidal radius of the dSph ($r_t = 7$ kpc was used
for these calculations) \citep{san07}, the main contribution to
$J(\Delta\Omega)$ comes from the regions $r < r_s \ll r_t$ and
therefore the choice of $r_t$ negligibly affects the $J$ value.
The main uncertainty for $J$ computation is due to the choice of
$\rho_s$ and $r_s$.  For Draco and Ursa Minor, $\rho_s$ and $r_s$ are
taken as the midpoints of the range from \citet{str07}.  For Willman
1, $\rho_s$ and $r_s$ are adopted from \citet{bri09b}.  The $J$ value Bo\"{o}tes 1
was calculated by Martinez and Bullock as discussed in section~\ref{sec:targets}.
The summary of the $J$ values calculated for each object is given in
Table~\ref{tbl:J}.

An estimated value of $J$ of order 10 is representative for all
observed dSphs, which is three orders of magnitude below the value
needed to constrain generic WIMP models with $m_\chi \gtrsim 100$ GeV/c$^2$.
Figure~\ref{fig:limit-plot} shows the exclusion region in the
$(m_\chi, \langle\sigma v\rangle)$ parameter space due to the
observations reported in this paper.  MSSM models shown in the figure
were produced with a random scan of the 7-parameter phase space
defined in the DarkSUSY package \citep{gondolo04} with the additional
WMAP \citep{spe07} constraint on the cosmological DM energy density.

Several astrophysical factors can increase the value of $J$ as compared
to the conservative estimates given in Table \ref{tbl:J}.  First, the
inner asymptotic behavior of the DM density may be steeper than
$\propto r^{-1}$ predicted by the NFW profile due to unaccounted
physical processes at small spatial scales.  The extreme assumption
would be the Moore profile \citep{moore99} $\propto r^{-1.5}$
asymptotically which generates a logarithmically divergent
self-annihilation flux indicating that another physical process, for
example self-annihilation, would limit the DM density in the central
regions of the galaxy.  A second factor that would increase the value of $J$
is deviations of the
DM distribution from a smooth average profile (substructures).
CDM N-body simulations predict substructures in DM halos
\citep{silk93,diemand05,diemand07} and the
effects on the DM self-annihilation have been
studied in these simulations.  In general any
regions of DM overdensity will enhance the self-annihilation flux; the
cumulative effect of these enhancements is usually referred to as the
boost factor.  \citet{str07} find a maximum boost factor of order
$10^2$ while a more detailed calculation that accounts for the
particle properties of the neutralino during formation of DM halos
suggests boost factors of order 10 and below \citep{martinez09}.  Thus,
present generation IACTs could be as close as two orders of magnitude in
sensitivity from constraining generic MSSM models.

Two effects related to the properties of the WIMP particle may improve
the chances of the detection of neutralino self-annihilation by
ground-based gamma-ray observatories.  Internal bremsstrahlung
gamma rays produced in neutralino self-annihilation recently
calculated by \citet{bri08} can significantly enhance dN/dE at the
energies comparable to $m_\chi$ for some MSSM models due to the
absence of the helicity suppression factor.  Effectively this
increases the value of the integral in Equation \ref{eqn:limit},
especially for the higher mass neutralino models.  In addition, the
$\left<\sigma v\right>$ for self-annihilation at the present
cosmological time may be considerably larger than at the time of
WIMP decoupling due to a velocity-dependent term in the cross-section and
the reduction of the kinetic energy of the WIMP due to the
cosmological expansion of the universe \citep{rob09,pie09}.

\section{Conclusions}     \label{sec:conclu}

We have carried out a search for VHE gamma rays from four dSphs:
Draco, Ursa Minor, Bo\"{o}tes 1, and Willman 1, as part of an indirect DM
search program at the VERITAS observatory.  The dSphs were selected
for proximity to Earth and for favorable estimates of the J factor
based on stellar kinematics data.  No significant gamma-ray excess was
observed from the four dSphs, and the derived upper limits on the
gamma-ray flux constrain the $\langle\sigma v \rangle$ for neutralino
pair annihilation as a function of neutralino mass to be $\lesssim
10^{-23}$ cm$^3$ s$^{-1}$ for $m_\chi \gtrsim 300$ GeV/c$^2$.  The obtained
$\langle\sigma v \rangle$ limits are three orders of magnitude above
generic predictions for MSSM models assuming an NFW DM density
profile, no boost factor, and no additional particle-related gamma-ray
flux enhancement factors.  Should the neglected effects be included,
the constraints on $\langle\sigma v \rangle$ in the most optimistic
regime could be pushed to $\lesssim 10^{-25}$ cm$^3$ s$^{-1}$.

To begin confronting the predictions of generic MSSM models through
observation of presently known dSphs, future ground-based observatories
will need a sensitivity at least an order of magnitude better than
present-day instruments.  The list of dSphs favorable for observations of DM
self-annihilation has grown over the last years by a factor of
roughly two, and it is anticipated that newly discovered dSphs may offer a
larger factor J.  The ongoing sky survey conducted by the Fermi
Gamma-ray Space Telescope (FGST) may also identify nearby higher DM
density substructures within the MW galaxy which could be followed up
by the IACT observatories.  Typical current exposures accumulated on
dSphs by IACTs are of order 20 hours, and ongoing observing programs
could feasibly increase the depth of these observations by a factor of
10 (a sensitivity increase of $\sim$3).  Improvements in background
rejection are anticipated to increase sensitivity by an additional
20-50\%.  The soon-to-be-operational upgrades, MAGIC-II and HESS-II,
as well as a planned VERITAS upgrade
will reduce the energy threshold and consequently increase the $dN/dE$
contribution by a factor as large as 10 thus providing an additional
sensitivity improvement.  With all these factors combined, the
$\langle\sigma v \rangle$ limits for $m_\chi \gtrsim 300$ GeV/c$^2$ will
begin to rule out the most favorable MSSM models assuming a moderate
boost factor.  Next generation IACT arrays now being planned such as
the Advanced Gamma-ray Imaging System (AGIS)
\footnote{http://www.agis-observatory.org/} and the Cherenkov
Telescope Array (CTA) \footnote{http://www.cta-observatory.org/} will
provide an order of magnitude increase in sensitivity and lower the
energy threshold by factor of $\sim$2 as compared to VERITAS.  These
instruments will be able to probe the bulk of the parameter space for
generic MSSM models with $m_\chi \simeq 300$ GeV/c$^2$ without strong
assumptions regarding potential flux enhancement factors.

This research is supported by grants from the US National Science
Foundation, the US Department of Energy, and the Smithsonian
Institution, by NSERC in Canada, by Science Foundation Ireland, and by
STFC in the UK.  We acknowledge the excellent work of the technical
support staff at the FLWO and the collaborating institutions in the
construction and operation of the instrument.
V.V.V. acknowledges the support of the U.S. National
Science Foundation under CAREER program (Grant No. 0422093).




\clearpage


\end{document}